\documentstyle[twoside,epsf]{article}

\catcode`\@=11
\long\def\@makefntext#1{
\protect\noindent \hbox to 3.2pt {\hskip-.9pt  
$^{{\eightrm\@thefnmark}}$\hfil}#1\hfill}		

\def\@makefnmark{\hbox to 0pt{$^{\@thefnmark}$\hss}}	
	
\def\ps@myheadings{\let\@mkboth\@gobbletwo
\def\@oddhead{\hbox{}
\rightmark\hfil\eightrm\thepage}   
\def\@oddfoot{}\def\@evenhead{\eightrm\thepage\hfil
\leftmark\hbox{}}\def\@evenfoot{}
\def\sectionmark##1{}\def\subsectionmark##1{}}

\oddsidemargin=\evensidemargin
\newcounter{sectionc}\newcounter{subsectionc}\newcounter{subsubsectionc}
\renewcommand{\section}[1] {\vspace{12pt}\addtocounter{sectionc}{1} 
\setcounter{subsectionc}{0}\setcounter{subsubsectionc}{0}\noindent 
	{\tenbf\thesectionc. #1}\par\vspace{5pt}}
\renewcommand{\subsection}[1] {\vspace{12pt}\addtocounter{subsectionc}{1} 
	\setcounter{subsubsectionc}{0}\noindent 
	{\bf\thesectionc.\thesubsectionc. {\kern1pt \bfit #1}}\par\vspace{5pt}}
\renewcommand{\subsubsection}[1] {\vspace{12pt}\addtocounter{subsubsectionc}{1}
	\noindent{\tenrm\thesectionc.\thesubsectionc.\thesubsubsectionc.
	{\kern1pt \tenit #1}}\par\vspace{5pt}}
\newcommand{\nonumsection}[1] {\vspace{12pt}\noindent{\tenbf #1}
	\par\vspace{5pt}}

\newcounter{appendixc}
\newcounter{subappendixc}[appendixc]
\newcounter{subsubappendixc}[subappendixc]
\renewcommand{\thesubappendixc}{\Alph{appendixc}.\arabic{subappendixc}}
\renewcommand{\thesubsubappendixc}
	{\Alph{appendixc}.\arabic{subappendixc}.\arabic{subsubappendixc}}

\renewcommand{\appendix}[1] {\vspace{12pt}
        \refstepcounter{appendixc}
        \setcounter{figure}{0}
        \setcounter{table}{0}
        \setcounter{lemma}{0}
        \setcounter{theorem}{0}
        \setcounter{corollary}{0}
        \setcounter{definition}{0}
        \setcounter{equation}{0}
        \renewcommand{\thefigure}{\Alph{appendixc}.\arabic{figure}}
        \renewcommand{\thetable}{\Alph{appendixc}.\arabic{table}}
        \renewcommand{\theappendixc}{\Alph{appendixc}}
        \renewcommand{\thelemma}{\Alph{appendixc}.\arabic{lemma}}
        \renewcommand{\thetheorem}{\Alph{appendixc}.\arabic{theorem}}
        \renewcommand{\thedefinition}{\Alph{appendixc}.\arabic{definition}}
        \renewcommand{\thecorollary}{\Alph{appendixc}.\arabic{corollary}}
        \renewcommand{\theequation}{\Alph{appendixc}.\arabic{equation}}
        \noindent{\tenbf Appendix \theappendixc #1}\par\vspace{5pt}}
\newcommand{\subappendix}[1] {\vspace{12pt}
        \refstepcounter{subappendixc}
        \noindent{\bf Appendix \thesubappendixc. {\kern1pt \bfit #1}}
	\par\vspace{5pt}}
\newcommand{\subsubappendix}[1] {\vspace{12pt}
        \refstepcounter{subsubappendixc}
        \noindent{\rm Appendix \thesubsubappendixc. {\kern1pt \tenit #1}}
	\par\vspace{5pt}}

\newcommand{\textlineskip}{\baselineskip=13pt}
\newcommand{\smalllineskip}{\baselineskip=10pt}

\def\eightcirc{
\begin{picture}(0,0)
\put(4.4,1.8){\circle{6.5}}
\end{picture}}
\def\eightcopyright{\eightcirc\kern2.7pt\hbox{\eightrm c}} 

\newcommand{\copyrightheading}[1]
	{\vspace*{-2.5cm}\smalllineskip{\flushleft
	{\footnotesize International Journal of Modern Physics B, #1}\\
	{\footnotesize $\eightcopyright$\, World Scientific Publishing
	 Company}\\
	 }}

\newcommand{\pub}[1]{{\begin{center}\footnotesize\smalllineskip 
	Received #1\\
	\end{center}
	}}

\def\abstracts#1#2#3{{
	\centering{\begin{minipage}{4.5in}\baselineskip=10pt\footnotesize
	\parindent=0pt #1\par 
	\parindent=15pt #2\par
	\parindent=15pt #3
	\end{minipage}}\par}} 

\renewenvironment{thebibliography}[1]			
	{\frenchspacing
	 \ninerm\baselineskip=11pt
	 \begin{list}{\arabic{enumi}.}
	{\usecounter{enumi}\setlength{\parsep}{0pt}
	 \setlength{\leftmargin 12.7pt}{\rightmargin 0pt} 
	 \setlength{\itemsep}{0pt} \settowidth
	{\labelwidth}{#1.}\sloppy}}{\end{list}}

\newcounter{itemlistc}
\newcounter{romanlistc}
\newcounter{alphlistc}
\newcounter{arabiclistc}

\newcommand{\fcaption}[1]{
        \refstepcounter{figure}
        \setbox\@tempboxa = \hbox{\footnotesize Fig.~\thefigure. #1}
        \ifdim \wd\@tempboxa > 5in
           {\begin{center}
        \parbox{5in}{\footnotesize\smalllineskip Fig.~\thefigure. #1}
            \end{center}}
        \else
             {\begin{center}
             {\footnotesize Fig.~\thefigure. #1}
              \end{center}}
        \fi}

\newcommand{\tcaption}[1]{
        \refstepcounter{table}
        \setbox\@tempboxa = \hbox{\footnotesize Table~\thetable. #1}
        \ifdim \wd\@tempboxa > 5in
           {\begin{center}
        \parbox{5in}{\footnotesize\smalllineskip Table~\thetable. #1}
            \end{center}}
        \else
             {\begin{center}
             {\footnotesize Table~\thetable. #1}
              \end{center}}
        \fi}

\def\@citex[#1]#2{\if@filesw\immediate\write\@auxout
	{\string\citation{#2}}\fi
\def\@citea{}\@cite{\@for\@citeb:=#2\do
	{\@citea\def\@citea{,}\@ifundefined
	{b@\@citeb}{{\bf ?}\@warning
	{Citation `\@citeb' on page \thepage \space undefined}}
	{\csname b@\@citeb\endcsname}}}{#1}}

\newif\if@cghi
\def\cite{\@cghitrue\@ifnextchar [{\@tempswatrue
	\@citex}{\@tempswafalse\@citex[]}}
\def\citelow{\@cghifalse\@ifnextchar [{\@tempswatrue
	\@citex}{\@tempswafalse\@citex[]}}
\def\@cite#1#2{{$\null^{#1}$\if@tempswa\typeout
	{IJCGA warning: optional citation argument 
	ignored: `#2'} \fi}}

\def\pmb#1{\setbox0=\hbox{#1}
	\kern-.025em\copy0\kern-\wd0
	\kern.05em\copy0\kern-\wd0
	\kern-.025em\raise.0433em\box0}

\def\fnt#1#2{\footnotetext{\kern-.3em
	{$^{\mbox{\scriptsize #1}}$}{#2}}}

\def\fpage#1{\begingroup
\voffset=.3in
\thispagestyle{empty}\begin{table}[b]\centerline{\footnotesize #1}
	\end{table}\endgroup}

\def\runninghead#1#2{\pagestyle{myheadings}
\markboth{{\protect\footnotesize\it{\quad #1}}\hfill}
{\hfill{\protect\footnotesize\it{#2\quad}}}}
\font\tenrm=cmr10
\font\tenit=cmti10 
\font\tenbf=cmbx10
\font\bfit=cmbxti10 at 10pt
\font\ninerm=cmr9

\font\eightrm=cmr8

\def\qed{\hbox{${\vcenter{\vbox{			
   \hrule height 0.4pt\hbox{\vrule width 0.4pt height 6pt
   \kern5pt\vrule width 0.4pt}\hrule height 0.4pt}}}$}}

\def\bsc{{\sc a\kern-6.4pt\sc a\kern-6.4pt\sc a}}	
\def\bflatex{\bf L\kern-.30em\raise.3ex\hbox{\bsc}\kern-.14em 
T\kern-.1667em\lower.7ex\hbox{E}\kern-.125em X} 

\begin{document}

\runninghead{
Tom Busche and Peter Kopietz}
{
How does a quadratic term in the energy dispersion 
modify... }

\normalsize\textlineskip
\thispagestyle{empty}
\setcounter{page}{1}

\copyrightheading{}			

\vspace*{0.88truein}

\fpage{1}
\centerline{\bf HOW DOES A QUADRATIC TERM IN THE ENERGY DISPERSION
MODIFY THE}
\vspace*{0.035truein}
\centerline{\bf SINGLE-PARTICLE GREEN'S FUNCTION OF THE 
TOMONAGA-LUTTINGER  MODEL?}
\vspace*{0.37truein}
\centerline{\footnotesize TOM BUSCHE AND PETER KOPIETZ}
\vspace*{0.015truein}
\centerline{\footnotesize\it 
Institut f\"{u}r Theoretische Physik 
der Universit\"{a}t G\"{o}ttingen, Bunsenstrasse 9,}
\baselineskip=10pt
\centerline{
\footnotesize\it 
D-37073 G\"{o}ttingen, Germany}

\vspace*{10pt}
\pub{November 1999}

\vspace*{0.21truein}
\abstracts{
We calculate the effect of a quadratic term
in the energy dispersion on the low-energy
behavior of the Green's function of
the spinless Tomonaga-Luttinger model (TLM).
Assuming that for small wave-vectors $q = k -  k_F$ 
the fermionic excitation energy 
relative to the Fermi energy
is $ v_F q + q^2 / (2m)$, 
we explicitly calculate the single-particle Green's function for finite but
small values of $ \lambda = q_c /(2k_F)$.
Here $k_F$ is the Fermi wave-vector, $q_c$ is
the maximal momentum transfered
by the interaction, and $v_F = k_F / m$ is the Fermi velocity.
Assuming equal forward scattering
couplings $g_2 = g_4$, we find that the dominant effect 
of the quadratic term in the energy dispersion 
is a renormalization
of the anomalous dimension.
In particular, at weak coupling 
the anomalous dimension is
$\tilde{\gamma} =
\gamma ( 1 - 2 \lambda^2 \gamma )$, 
where $\gamma$ is the anomalous dimension of the TLM.
We also show how to treat the change of the chemical
potential due to the interactions within the
functional bosonization approach in arbitrary dimensions.
}{}{}

\vspace*{1pt}\textlineskip	
\section{Introduction}	
\vspace*{-0.5pt}
\noindent
The exactly solvable Tomonaga-Luttinger model (TLM)
has been extremely useful to gain a better understanding
of the electron-electron interactions 
in the normal metallic state of
electrons in one spatial dimension 
($d=1$)\cite{Tomonaga50,Luttinger63,Mattis65,Solyom79,Haldane81,Stone94}.
In this model  only the forward scattering part
of the electron-electron interaction 
is retained, which
in $d=1$ can be parameterized by two phenomenological
coupling constants
$g_2 $ and $g_4 $(Ref.\cite{Solyom79}). 
The second important feature of the TLM, which is crucial
for its exact solubility via
bosonization\cite{Mattis65,Haldane81,Stone94}
or other methods\cite{Dzyaloshinskii73,Bohr81}, is the 
fact that the energy dispersion of the electrons 
is exactly linear for all wave-vectors.
Because the low-energy and long-wavelength physics 
is determined by the degrees 
of freedom in the vicinity of the Fermi points,
it is reasonable to expect that the replacement
of a general energy dispersion by 
the linear term in the expansion close to the
Fermi points is justified.

The single-particle Green's function
$G ( x , t )$ of the TLM in the space-time domain
has been calculated by many 
authors\cite{Mattis65,Haldane81,Stone94,Dzyaloshinskii73,Bohr81}. 
A detailed discussion of the rather  complicated 
behavior of the Fourier transform  $G ( k , \omega )$
of $G ( x , t )$ in the momentum-frequency domain can be 
found in Ref.\cite{Meden92}.  
The structure of $G ( k , \omega )$ is 
fundamentally different from the well-known
behavior of the Green's function of a Fermi liquid:
There is no quasi-particle peak, but 
$G ( k , \omega )$ contains instead 
algebraic singularities, which can be interpreted physically
in terms of bosonic charge- and spin- excitations 
propagating with different velocities. 
The absence of a quasiparticle peak implies also that
the density of states of the TLM vanishes at the Fermi energy,
and that the momentum distribution function exhibits
only an algebraic singularity at the Fermi surface, in contrast
to the jump discontinuity of a Fermi liquid.

Haldane\cite{Haldane81} has emphasized 
that the above features are not special to the TLM, but
are generic to the normal metallic state of interacting
electrons in $d=1$.  He proposed
to call such a state a Luttinger liquid, in contrast
to the well-known Fermi liquid, which is the
normal metallic state of interacting electrons in $d=3$.
In the seminal work\cite{Haldane81} Haldane also 
emphasized that the
various deviations from the idealized
TLM can in principle be treated  perturbatively 
within the framework of bosonization,
and should only lead to a re-definition of the
phenomenological parameters that appear in the 
Hamiltonian of the TLM, namely the 
Fermi velocity $v_F$ and the forward scattering
couplings $g_2$ and $g_4$.
For example, in bosonic language
the non-linear terms in the expansion
of the energy dispersion close to the Fermi points
give rise to interactions between the Tomonaga-Luttinger bosons,
which can be treated perturbatively
within conventional bosonic many-body theory\cite{Fetter71}.
In Ref.\cite{Haldane81} Haldane performed an
expansion of the equal-time Green's function 
to lowest order in the curvature parameter $1/m$,
and pointed out 
that the non-linear terms in the energy dispersion
can be expected to modify the anomalous dimension of the TLM.
However, the precise value of this renormalization
was not obtained in Ref.\cite{Haldane81}.
This will be done in the present work with the help 
of  the  functional integral formulation of 
bosonization\cite{Kopietz97,Kopietz96,Kopietz96b}.

\section{
Interacting electrons
with dominant forward scattering}
\label{sec:Interacting}
\noindent
In this section we 
shall briefly describe how 
the single-particle Green's function\cite{footnote1} 
$G(\bf{k},\omega)$ of interacting fermions
with dominant forward scattering 
can be calculated within the
functional bosonization formalism\cite{Kopietz97,Kopietz96,Kopietz96b}
in arbitrary dimensions.
In particular, we shall rely on the results of 
Refs.\cite{Kopietz97,Kopietz96b}, where a systematic 
method has been developed
to handle the non-linear terms in the expansion of 
the energy dispersion close to the Fermi surface.
It turns out for our calculation 
it is important to take the renormalization of the chemical
potential due to the interaction into account. This 
renormalization has not been discussed in 
previous works\cite{Kopietz97,Kopietz96,Kopietz96b,Houghton94,CastroNeto94}.
Although we are ultimately interested in $d=1$,
all equations in this section are valid in arbitrary $d$.

\subsection{
Calculation of the single-particle Green's function
via functional bosonization}
\noindent
For simplicity, we shall ignore the spin degree of freedom.
We also assume
that
the bare energy dispersion is
$\epsilon_{\bf{k}} = \frac{ {\bf{k}}^2}{2m}$, and that the
Fourier transform $f_{\bf{q}}$ of the bare interaction
between two electrons with momenta
${\bf{k}}$ and ${\bf{k}}^{\prime}$ depends
only on the absolute value of the momentum
transfer ${\bf{k}} - {\bf{k}}^{\prime}$.

\begin{figure}
\epsfysize4cm 
\hspace{23mm}
\epsfbox{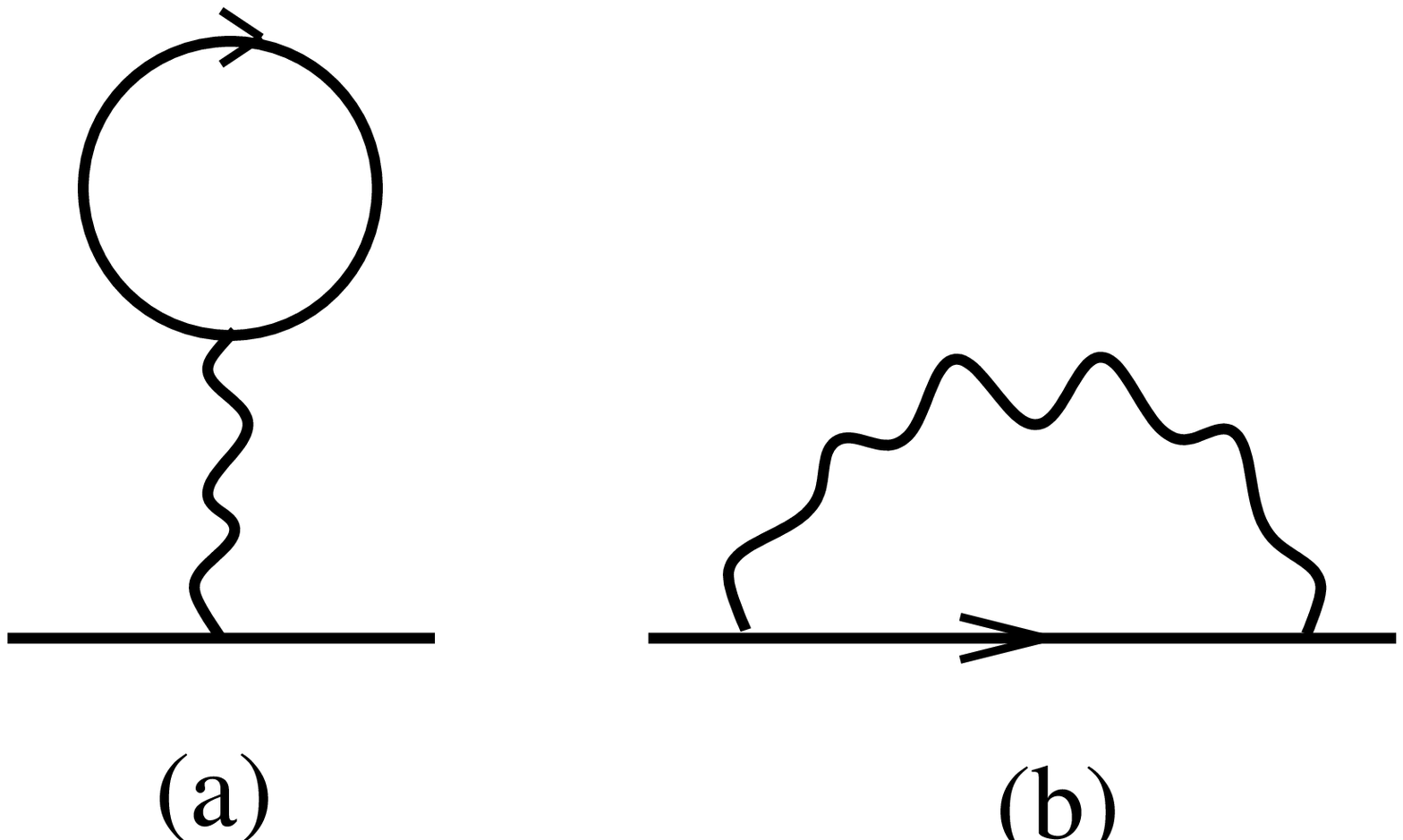}
\fcaption{
Feynman diagrams which renormalize
the chemical potential.
Here the wavy-line is the screened interaction and the
arrows represent the Green's function.
(a) ''Tadpole''-diagram; to first  order in the interaction
this is just the Hartree
correction to the self-energy.
(b) ''Sunrise''-diagram; to first order this is the Fock correction.
}
\label{fig:chempot}
\end{figure}

It is convenient to consider the Matsubara Green's function
$G ( {\bf{k}} , i \tilde{\omega}_n )$,
where $\tilde{\omega}_n = 2 \pi ( n + \frac{1}{2} ) / \beta$
are fermionic Matsubara frequencies. 
The inverse temperature is denoted by $\beta$.
Suppose that we are interested in
$G ( {\bf{k}} , i \tilde{\omega}_n )$ at some momentum
${\bf{k}}$ in the vicinity of the Fermi surface.
It is then convenient to  measure 
all momenta relative to some reference point ${\bf{k}}^{\alpha}$
that is close to the Fermi surface, such  that
the length of
${\bf{q}} = {\bf{k}} - {\bf{k}}^{\alpha} $
is small compared with $| {\bf{k}}^{\alpha} |$.
Let  us therefore define 
 \begin{equation}
 G^{\alpha} ( {\bf{q}} , i \tilde{\omega}_n ) = 
 G ( {\bf{k}}^{\alpha} + {\bf{q}} , i \tilde{\omega}_n )
 \; ,
 \label{eq:Gdef}
 \end{equation}
where the index $\alpha$ indicates that ${\bf{q}}$ is measured
relative to the reference  point ${\bf{k}}^{\alpha}$
in the vicinity of the Fermi surface.
The precise choice of ${\bf{k}}^{\alpha}$
will  be discussed shortly.
Note that the Fermi surface is defined as the surface in momentum space
where the momentum distribution
 \begin{equation}
 n ( {\bf{k}} ) = \frac{1}{\beta} \sum_n 
 G ( {\bf{k}} , i \tilde{\omega }_n ) 
 \label{eq:momdis}
 \end{equation}
exhibits some kind of non-analytic behavior.
Thus,  in general we have to solve the many-body problem
in order to know the precise location of the Fermi surface.
Of course, 
{\it{if}} Luttinger's theorem\cite{Luttinger60,Blagoev97}
turns out to be valid in our system, then 
the volume of the Fermi surface is not changed  
when the interaction is switched on at constant density.
For a spherically symmetric Fermi surface
this implies that its  
radius is precisely identical with the
Fermi wave-vector $k_F$ of the non-interacting system,
 \begin{equation}
 k_F = \sqrt{ 2 m \mu_0} 
 \; ,
 \label{eq:kfdef}
 \end{equation}
where $\mu_0$ is the chemical potential of the
system without interactions at the same density.
In general, however, the precise location of the
Fermi surface is a priori not known. 
In fact, we cannot even exclude the possibility
that for some type of interaction the momentum distribution 
is an analytic function of ${\bf{k}}$, in which case
the  system does not have a Fermi surface.
Note that
in the shifted coordinate system 
the energy dispersion is
 \begin{equation}
 \epsilon_{ {\bf{k}}^{\alpha} + {\bf{q}} }
 =  \epsilon_{ {\bf{k}}^{\alpha} } + 
 {\bf{v}}^{\alpha} \cdot {\bf{q}}  + \frac{ {\bf{q}}^2}{2 m}
 \; ,
 \label{eq:epsexp}
 \end{equation}
where ${\bf{v}}^{\alpha} = {\bf{k}}^{\alpha} / m$. 
Assuming that the system is confined to a finite volume $V$, 
the real-space imaginary-time 
Fourier transform of $G^{\alpha} ( {\bf{q}} , i \tilde{\omega}_n )$ is
 \begin{equation}
 G^{\alpha} ( {\bf{r}} , \tau )
 = 
 \frac{1}{\beta V } \sum_{ {\bf{q}}  n} 
 e^{ i (  {\bf{q}} \cdot {\bf{r}} - \tilde{\omega}_n \tau )}
 G^{\alpha} ( {\bf{q}} , i \tilde{\omega}_n )
 \label{eq:GFTdef}
 \; .
 \end{equation}
After the usual Hubbard-Stratonovich transformation\cite{Kopietz97}
 $G^{\alpha} ( {\bf{r}} , \tau )$ can 
be written as
 \begin{equation}
 G^{\alpha} ( {\bf{r}} - {\bf{r}}^{\prime} , \tau - \tau^{\prime} )
 = 
 \langle
 {\cal{G}} ^{\alpha} ( {\bf{r}} , {\bf{r}}^{\prime} , \tau , \tau^{\prime} )
 \rangle
 \label{eq:Gav}
 \; ,
 \end{equation}
where 
 \begin{equation}
 \left[ - \partial_{\tau}
 - {\bf{v}}^{\alpha} \cdot \hat{\bf{P}}_{\bf{r}} -
 \frac{ \hat{\bf{P}}_{\bf{r}}^2 }{2 m} 
 - \epsilon_{  {\bf{k}}^{\alpha} } + \mu
 - i \phi ( {\bf{r}} , \tau )
 \right]
 {\cal{G}} ^{\alpha} ( {\bf{r}} , {\bf{r}}^{\prime} , \tau , \tau^{\prime} )
 = \delta ( {\bf{r}} - {\bf{r}}^{\prime} )
 \delta^{\ast} ( \tau - \tau^{\prime} )
 \label{eq:Gdif}
 \; .
 \end{equation}
Here $\hat{\bf{P}}_{\bf{r}} = -i \nabla_{\bf{r}}$ is the momentum operator,
$\delta^{\ast} ( \tau ) = \frac{1}{\beta} \sum_n e^{i \tilde{\omega}_n \tau}$
is the antiperiodic $\delta$-function, and
$\langle \ldots \rangle$ denotes functional averaging
with respect to the effective action of the Hubbard-Stratonovich
field $\phi ( {\bf{r}} , \tau )$. The chemical potential of the interacting system is denoted by
$\mu$. 
Let us now choose the reference point ${\bf{k}}^{\alpha}$ such that
 \begin{equation}
 \epsilon_{ {\bf{k}}^{\alpha} }
 = \mu  - \langle i \phi ( {\bf{r}} , \tau )
 \rangle
 \; .
 \label{eq:kalphadef}
 \end{equation}
Diagrammatically the term
 $ \langle i \phi ( {\bf{r}} , \tau ) \rangle $ 
 corresponds to the sum of all tadpole diagrams which
renormalize the chemical potential, see  
Fig.\ref{fig:chempot}(a).
The leading diagram of this type is just the 
usual Hartree diagram.
We shall include the sum of these diagrams into the definition
of the zeroth order Green's function, 
 \begin{eqnarray}
 G_0^{\alpha} ( {\bf{q}} , i \tilde{\omega}_n )
 & = &
 \frac{1}{ i \tilde{\omega}_n - \epsilon_{ {\bf{k}}^{\alpha} +
 {\bf{q}} } + \mu - \langle i \phi ( {\bf{r}} , \tau ) \rangle }
 \nonumber
 \\
 & = & \frac{1}{
  i \tilde{\omega}_{n} -  {\bf{v}}^{\alpha} \cdot {\bf{q}} 
  - \frac{ {\bf{q}}^2}{2m} }
  \; .
  \label{eq:G0def}
  \end{eqnarray}
The effective action  of the Hubbard-Stratonovich field
is given by
 \begin{equation}
 S_{\rm eff} \{ \phi  \} = \frac{\beta V }{2  } 
 \sum_ { {\bf{q}} n }
 f_{\bf{q}}^{-1}
 {\phi}_{ - {\bf{q}},  - n } 
 {\phi}_{ {\bf{q}}, n } 
 - {\rm Tr} \ln [ 1 - 
 i \hat{G}_0 \hat{\phi} ]
 \; ,
 \label{eq:Seffdef}
 \end{equation}
where the trace
is over momentum-frequency space, with the infinite matrices
$\hat{G}_0$ and $\hat{\phi}$ given by
 \begin{equation}
 [ \hat{G}_0 ]_{ {\bf{q}} n , {\bf{q}}^{\prime} n^{\prime} }
  = 
  \delta_{ {\bf{q}} , {\bf{q}}^{\prime} }
 \delta_{n , n^{\prime}}
 G_0^{\alpha} ( {\bf{q}} , i \tilde{\omega}_n )
 \; ,
 \label{eq:G0matdef}
 \end{equation}
 \begin{equation}
 [ \hat{\phi} ]_{ {\bf{q}} n , {\bf{q}}^{\prime} n^{\prime} }
  =   {\phi}_{ {\bf{q}} - {\bf{q}}^{\prime} ,
 n - n^{\prime} }
 \; .
 \label{eq:phimatdef}
 \end{equation}
Here
 \begin{equation}
 {\phi}_{ {\bf{q}} , n }  =
 \frac{1}{\beta V } \int d {\bf{r}} \int_{- \beta /2}^{\beta /2} d \tau 
 e^{ - i ( {\bf{q}} \cdot {\bf{r}} - \omega_n \tau )}
  [  \phi ( {\bf{r}} , \tau ) 
  - \langle 
    \phi ( {\bf{r}} , \tau )  \rangle ]
 \label{eq:phiftdef}
 \; ,
 \end{equation}
where $\omega_n = 2 \pi n / \beta$ are 
bosonic Matsubara frequencies.
In Eq.(\ref{eq:phiftdef}) we have
subtracted the
average $\langle \phi ( {\bf{r}} , \tau ) \rangle$, which
by translational invariance is independent
of ${\bf{r}}$ and $\tau$. This
subtraction modifies only the 
zeroth Fourier component
$\phi_0 \equiv \phi_{  {\bf{q}} = 0 , n = 0}$
such that 
$\langle \phi_0 \rangle = 0$.
Note that 
for non-zero ${\bf{q}}$ or $\omega_n$ 
translational invariance implies
$\langle {\phi}_{ {\bf{q}}  n }  \rangle = 0$.

From Eqs.(\ref{eq:Gdef}) and (\ref{eq:GFTdef})
it is clear that the Fourier transform $G ( {\bf{r}} , \tau )$ of the
physical Matsubara Green's function $G ( {\bf{k}} , i \tilde{\omega}_n)$
is simply
 \begin{equation}
 G ( {\bf{r}} , \tau ) = e^{i {\bf{k}}^{\alpha} \cdot {\bf{r}} }
 G^{\alpha} ( {\bf{r}} , \tau )
 \; .
 \label{eq:Gtotal}
 \end{equation}
If one works with locally linearized energy dispersion
the Fermi surface has to be subdivided into several
sufficiently small sectors\cite{Kopietz97,Houghton94},
which are labelled by the discrete index $\alpha$.
In this case 
the right-hand side of Eq.(\ref{eq:Gtotal}) should
be summed over all these sectors in order to recover the
physical Green's function. 
Note that the sectorization would require the introduction
of cutoffs in momentum space.
However, for our purpose such a subdivision
into sectors is really not necessary {\it{as long as we
restrict ourselves to the calculation of
$G ( {\bf{k}}^{\alpha} + {\bf{q}} , \omega )$
for $| {\bf{q}} | \ll | {\bf{k}}^{\alpha} |$.}}
Of course, without sectorization the curvature of the
Fermi surface has to be taken into account\cite{Kopietz97,Kopietz96b}.

It turns out that for interactions that involve only
small momentum transfers $ | {\bf{q}}  | \ll | {\bf{k}}^{\alpha}|$
it is sufficient to perform the averaging in Eq.(\ref{eq:Gav}) within
the Gaussian approximation\cite{Kopietz97,Kopietz96}, which amounts
to an expansion of the logarithm in Eq.(\ref{eq:Seffdef})
to the second order. Within this approximation
the effective action 
is given by
 \begin{equation}
 S_{{\rm eff}} \{ \phi  \} \approx  i \beta V \rho_0  \phi_0 + \frac{\beta V }{2  } 
 \sum_ { {\bf{q}} n }
 f_{\rm RPA}^{-1}  ( {\bf{q}} , i \omega_n ) 
 \phi_{ - {\bf{q}},  - n } 
 \phi_{ {\bf{q}}, n } 
 \; ,
 \label{eq:Seff2def}
 \end{equation}
where
 \begin{equation}
 \rho_0 = \frac{1}{\beta V} \sum_{ {\bf{q}} n }
 G_0^{\alpha} ( {\bf{q}} , i \tilde{\omega}_n )
 \label{eq:dens}
 \; , 
 \end{equation}
and 
 \begin{equation}
 f_{\rm RPA} ( {\bf{q}} , i \omega_n ) = 
 \frac{ f_{\bf{q}} }{ 1 + f_{\bf{q}} \Pi_0 ( {\bf{q}} , i \omega_n ) }
 \label{eq:frpadef}
 \end{equation}
 is the
effective interaction within the random-phase 
approximation (RPA).
The ``non-interacting'' polarization
can be written as
 \begin{equation}
 \Pi_0 ( {\bf{q}} , i \omega_n )
 = - \frac{1}{\beta V}
 \sum_{ {\bf{q}}^{\prime} n^{\prime} }
 G_0^{\alpha} ( {\bf{q}}^{\prime} , i \tilde{\omega}_{n^{\prime}} )
 G_0^{\alpha} ( {\bf{q}}^{\prime} + {\bf{q}} , i \tilde{\omega}_{n^{\prime}+n} )
 \; .
 \label{eq:Pol}
 \end{equation}

Let us now focus on the solution of Eq.(\ref{eq:Gdif}).
Making the
generalized Schwinger-ansatz\cite{Schwinger62,Kopietz97,Kopietz96b}
 \begin{equation}
 {\cal{G}}^{\alpha} ( {\bf{r}} , {\bf{r}}^{\prime} , \tau , \tau^{\prime} )
 =
 {\cal{G}}_1^{\alpha} ( {\bf{r}} , {\bf{r}}^{\prime} , \tau , \tau^{\prime} )
 e^{ 
 \Phi^{\alpha} ( {\bf{r}} , \tau )
 -
 \Phi^{\alpha} ( {\bf{r}}^{\prime} , \tau^{\prime} )
 }
 \label{eq:schwinger}
 \; ,
 \end{equation}
we obtain with ${\bf{k}}^{\alpha}$ given by
Eq.(\ref{eq:kalphadef})
 \begin{eqnarray}
 \left[ - \partial_{\tau}
 - {\bf{v}}^{\alpha} \cdot \hat{\bf{P}}_{\bf{r}} -
 \frac{ \hat{\bf{P}}_{\bf{r}}^2 }{2 m} - \frac{ {\bf{A}}^{\alpha} 
 ({\bf{r}} , \tau ) }{m}
 \cdot \hat{\bf{P}}_{\bf{r}} 
 \right]
 {\cal{G}}_1^{\alpha} ( {\bf{r}} , {\bf{r}}^{\prime} , \tau , \tau^{\prime} )
 + 
 {\cal{G}}_1^{\alpha} ( {\bf{r}} , {\bf{r}}^{\prime} , \tau , \tau^{\prime} )
 \nonumber
 \\
 & & \hspace{-115mm} \times
 \left\{
 \left[ - \partial_{\tau}
 - {\bf{v}}^{\alpha} \cdot \hat{\bf{P}}_{\bf{r}} -
 \frac{ \hat{\bf{P}}_{\bf{r}}^2 }{2 m}  \right] \Phi^{\alpha} ( {\bf{r}} , \tau )
 - i [ \phi( {\bf{r}} , \tau ) - \langle \phi ( {\bf{r}} , \tau )
 \rangle ] - 
 \frac{ [ {\bf{A}}^\alpha ({\bf{r}} , \tau )]^2 }{2m}
 \right\}
 \nonumber
 \\
 & & \hspace{-115mm}
 = \delta ( {\bf{r}} - {\bf{r}}^{\prime} )
 \delta^{\ast} ( \tau - \tau^{\prime} )
 \label{eq:Gsub}
 \; .
 \end{eqnarray}
Here 
 \begin{equation}
 {\bf{A}}^{\alpha} ( {\bf{r}} , \tau ) = \hat{\bf{P}}_{\bf{r}} 
 \Phi^{\alpha} ( {\bf{r}} , \tau )
 \end{equation}
is a longitudinal vector potential.
Eq.(\ref{eq:schwinger}) is a gauge
transformation.  Obviously, there are infinitely many
different choices for  the gauge factor
$\Phi^{\alpha} ( {\bf{r}} , \tau )$.
For example, if the bare interaction $f_{\bf{q}}$ is the
three-dimensional Coulomb interaction, then 
${\bf{A}}^{\alpha} ( {\bf{r}} , \tau )$ can be
identified with the longitudinal
part of the usual vector potential of electromagnetism
if we choose
$- \partial_{\tau} \Phi^{\alpha}  ( {\bf{r}} , \tau ) = i 
\phi ( {\bf{r}} , \tau )$. 
In Ref.\cite{Kopietz98} it was shown that this choice 
is useful to resum the leading logarithmic singularities
in the perturbative expansion of the 
Green's function of two-dimensional disordered electrons 
that interact with long-range Coulomb forces.
To include the  renormalization
of the chemical potential into 
the functional bosonization approach, a slight modification
of the gauge choice given in Refs.\cite{Kopietz97,Kopietz96b} is necessary.
It is easy to see that Eq.(\ref{eq:Gsub}) can be satisfied if 
 \begin{eqnarray}
 \left[ - \partial_{\tau}
 - {\bf{v}}^{\alpha} \cdot \hat{\bf{P}}_{\bf{r}} -
 \frac{ \hat{\bf{P}}_{\bf{r}}^2 }{2 m} - \frac{ {\bf{A}}^{\alpha} 
 ({\bf{r}} , \tau ) }{m}
 \cdot \hat{\bf{P}}_{\bf{r}} 
 - i \phi_0
 - D_0
 \right]
 {\cal{G}}_1^{\alpha} ( {\bf{r}} , {\bf{r}}^{\prime} , \tau , \tau^{\prime} )
 \nonumber
 \\
 & & \hspace{-80mm}
 = \delta ( {\bf{r}} - {\bf{r}}^{\prime} )
 \delta^{\ast} ( \tau - \tau^{\prime} )
 \; ,
 \end{eqnarray}
 \begin{eqnarray}
 \left[ - \partial_{\tau}
 - {\bf{v}}^{\alpha} \cdot \hat{\bf{P}}_{\bf{r}} -
 \frac{ \hat{\bf{P}}_{\bf{r}}^2 }{2 m}  \right] 
 \Phi^{\alpha} ( {\bf{r}} , \tau ) 
 \nonumber
 \\
 & & \hspace{-50mm} =
  i [ \phi( {\bf{r}} , \tau ) -  \langle \phi ( {\bf{r}} , \tau )
  \rangle   - \phi_0 ]
 + \frac{ [ \hat{\bf{P}}_{\bf{r}} 
 \Phi^\alpha ({\bf{r}} , \tau )]^2 }{2m} - D_0
 \label{eq:eik}
 \; .
 \end{eqnarray}
 Here $\phi_0$ is the
 zeroth Fourier component of $\phi ( {\bf{r}} , \tau )
 - \langle \phi ( {\bf{r}} , \tau ) \rangle$ (see
 Eq.(\ref{eq:phiftdef})),
and  $D_0$  the average diamagnetic energy,
 \begin{eqnarray}
 D_0 & = & 
 \frac{1}{\beta V}
 \int d {\bf{r}} \int_{- \beta / 2}^{\beta /2} d \tau
 \frac{ [ \hat{\bf{P}}_{\bf{r}} {{\Phi}}^\alpha ({\bf{r}} , \tau )]^2 }{2m}
 \nonumber
 \\
 & = & 
 \frac{1}{\beta V}
 \int d {\bf{r}} \int_{- \beta /2}^{\beta / 2} d \tau
 \frac{ [ {\bf{A}}^\alpha ({\bf{r}} , \tau )]^2 }{2m}
 \; .
 \label{eq:D0def}
 \end{eqnarray}
By construction the Fourier component corresponding to
$ {\bf{q}} = 0$ and $\omega_n = 0$ of the
right-hand side of Eq.(\ref{eq:eik}) vanishes, so that
in the iterative solution of the eikonal equation (\ref{eq:eik})
ambiguities due to vanishing eigenvalues of the
differential operator on the left-hand side are
avoided. Such an explicit separation of the
zeroth Fourier component, which was not discussed 
in Refs.\cite{Kopietz97,Kopietz96b}, is important to 
incorporate the renormalization of the chemical
potential into functional bosonization.

At this point we solve the eikonal equation (\ref{eq:eik}) 
to first order in $\phi ( {\bf{r}} , \tau )$, which is
sufficient to calculate the Debye-Waller factor
arising after the averaging  from the gauge-factor in Eq.(\ref{eq:schwinger})
to first order in the RPA interaction.
The functional averaging of the ''prefactor'' Green's function
is then performed within the self-consistent
Born approximation, as described in detail in Ref.\cite{Kopietz97}.
The final result for the Green's function of the interacting many-body system
for wave-vectors close to
${\bf{k}}^{\alpha}$ is
\cite{Kopietz97,Kopietz96b}
 \begin{equation}
 G ( {\bf{k}}^{\alpha} + {\bf{q}} , i
 \tilde{\omega}_n ) = \int d {\bf{r}} \int_{- \beta /2}^{\beta / 2} d \tau
 e^{ - i ( {\bf{q}} \cdot {\bf{r}}
  - \tilde{\omega}_n \tau ) } \tilde{G}^{\alpha} ( {\bf{r}} , \tau )
 e^{Q_1^{\alpha} ( {\bf{r}} , \tau )}
 \; ,
 \label{eq:Gfinal}
 \end{equation}
with the Debye-Waller factor given by
 \begin{equation}
 Q_1^{\alpha} ( {\bf{r}} , \tau )
 = \frac{1}{\beta V} \sum_{ {\bf{q}} n} 
 \frac{ 
 f_{\rm RPA} ( {\bf{q}} , i \omega_n )
 [ 1 - \cos ( {\bf{q}} \cdot {\bf{r}} - \omega_n \tau ) ]}{
 [ i \omega_n - {\bf{v}}^{\alpha} \cdot {\bf{q}}
 - \frac{ {\bf{q}}^2}{2m} ]
 [ i \omega_n - {\bf{v}}^{\alpha} \cdot {\bf{q}}
 + \frac{ {\bf{q}}^2}{2m} ]}
 \; ,
 \label{eq:DebyeWaller}
 \end{equation}
and the prefactor Green's function
 \begin{equation}
 \tilde{G}^{\alpha} ( {\bf{r}} , \tau )
  =  \frac{1}{\beta V}\sum_{ {\bf{q}}n }
 e^{i ( {\bf{q}} \cdot {\bf{r}} - \tilde{\omega}_n \tau )}
 [ 1 + Y^{\alpha} ( {\bf{q}} , i \tilde{\omega}_n ) ]
 G_1^{\alpha} ( {\bf{q}} , i \tilde{\omega}_n )
 \; .
 \label{eq:prefG}
 \end{equation}
Here
 \begin{equation}
 G_1^{\alpha} ( {\bf{q}} , i \tilde{\omega}_n ) =
 \frac{1}{
 i \tilde{\omega}_n - {\bf{v}}^{\alpha} \cdot {\bf{q}}
 - \frac{ {\bf{q}}^2}{2m} -  \delta \mu_F
 - \Sigma^{\alpha}_1 ( 
 {\bf{q}} , i \tilde{\omega}_n ) }
 \; ,
 \label{eq:G1def}
 \end{equation}
 \begin{eqnarray}
 \Sigma^{\alpha}_1 ( 
 {\bf{q}} , i \tilde{\omega}_n )   & = &
 - \frac{1}{\beta V} \sum_{ {\bf{q}}^{\prime} n^{\prime}
 } f_{ {\rm RPA} } ( {\bf{q}}^{\prime} , i
 \omega_{n^{\prime} }  ) G_1^{\alpha} (  {\bf{q }} + {\bf{q}}^{\prime} ,
 i \tilde{\omega}_{ n + n^{\prime} } )
 \nonumber
 \\
 &  &  \hspace{-15mm} \times
 \frac{ \frac{ ( {\bf{q}} \cdot {\bf{q}}^{\prime} ) {\bf{q}}^{\prime 2} }{m^2}
 +  ( \frac{ {\bf{q}} \cdot {\bf{q}}^{\prime}}{m} )^2 }{
 [ i \omega_{n^{\prime} } - {\bf{v}}^{\alpha} \cdot {\bf{q}}^{\prime}
 - \frac{ {\bf{q}}^{\prime 2}}{2m} ]
 [ i \omega_{n^{\prime}} - {\bf{v}}^{\alpha} \cdot {\bf{q}}^{\prime}
 + \frac{ {\bf{q}}^{\prime 2}}{2m} ]}
 \label{eq:sigma1}
 \; ,
 \end{eqnarray}
 \begin{eqnarray}
 Y^{\alpha} ( 
 {\bf{q}} , i \tilde{\omega}_n )   & = &
  \frac{1}{\beta V} \sum_{ {\bf{q}}^{\prime} n^{\prime}
 } f_{ {\rm RPA} } ( {\bf{q}}^{\prime} , i
 \omega_{n^{\prime} }  ) G_1^{\alpha} (  {\bf{q }} + {\bf{q}}^{\prime} ,
 i \tilde{\omega}_{ n + n^{\prime} } )
 \nonumber
 \\
 &  &  \hspace{-15mm} \times
 \frac{ \frac{ {\bf{q}}^{\prime 2}}{m}
 + 2 \frac{ {\bf{q}} \cdot {\bf{q}}^{\prime} }{m} }{
 [ i \omega_{n^{\prime} } - {\bf{v}}^{\alpha} \cdot {\bf{q}}^{\prime}
 - \frac{ {\bf{q}}^{\prime 2}}{2m} ]
 [ i \omega_{n^{\prime}} - {\bf{v}}^{\alpha} \cdot {\bf{q}}^{\prime}
 + \frac{ {\bf{q}}^{\prime 2}}{2m} ]}
 \label{eq:Y}
 \; ,
 \end{eqnarray}
and  
 \begin{eqnarray}
 \delta \mu_F 
 = &\langle D_0 \rangle &
  =  - \frac{1}{\beta V} \sum_{ {\bf{q}}  n }
 f_{\rm RPA} ( {\bf{q}} , i \omega_n )
 \nonumber
 \\
 &  & \times
 \frac{ \frac{ {\bf{q}}^2 }{2m} }
 {[ i \omega_n - {\bf{v}}^{\alpha} \cdot {\bf{q}}
 - \frac{ {\bf{q}}^2}{2m} ]
 [ i \omega_n - {\bf{v}}^{\alpha} \cdot {\bf{q}}
 + \frac{ {\bf{q}}^2}{2m} ]}
 \; .
 \label{eq:muF}
 \end{eqnarray}
Note that
$\delta \mu_F$ can be identified 
with the renormalization 
of the chemical potential arising
from the usual ''sunrise'' -diagram shown in Fig.\ref{fig:chempot}(b).
To leading order in the bare interaction, this is just 
the Fock diagram.
Note that the Hartree renormalization
of the chemical potential has already been taken into
account in the definition of the
zeroth order Green's function 
by a suitable choice of the reference point
${\bf{k}}^{\alpha}$, see Eq.(\ref{eq:kalphadef}).
This different treatment of the two renormalizations
of the chemical potential is justified
if we formally treat $1/m$ as a small parameter.
Obviously, $\delta \mu_F \propto 1/m$, whereas
for the Hartree correction that appears in Eq.(\ref{eq:kalphadef})
we obtain to leading order
 \begin{equation}
 \delta \mu_H =
 \langle i \phi_0 \rangle = \rho_0 f_{\rm RPA} ( 0,0 )
 \label{eq:muH}
 \; .
 \end{equation}
Keeping in mind that in $d$ dimensions
$\rho_0 \propto k_F^d \propto  m^d$, it 
is clear that the Hartree renormalization
of the chemical potential diverges if we let
$m \rightarrow \infty$ holding $v_F = k_F / m $ constant.
This is the reason why this renormalization cannot be
treated within a $1/m$-expansion, but should be
included in the definition of the
zeroth order Green's function\cite{KS}.

It is important to emphasize that 
the above expansion is quite different from the
expansion in powers of $1/m$ proposed by
Haldane\cite{Haldane81}. The advantage of our method
is that in the absence of interactions 
we recover the exact free Green's function
with non-linear energy dispersion, which
contains of course infinite orders in $1/m$.
In Sec.3 it will become clear
that the effective small parameter in our expansion
is  proportional to $f_{\rm RPA} / m$, so that
we obtain the exact result in both limits
$f_{\rm RPA} \rightarrow 0$ or $1/m \rightarrow 0$.

\subsection{
Luttinger's theorem}
\label{sec:chempot}
\noindent
The Fermi surface of an interacting Fermi system can be defined
as the surface in momentum space where the momentum distribution
defined in Eq.(\ref{eq:momdis})
has some kind of non-analyticity in the zero-temperature limit $\beta 
\rightarrow \infty$. 
The Fermi surface is in general not identical with the
surface defined by Eq.(\ref{eq:kalphadef}), so that
our reference point ${\bf{k}}^{\alpha}$ is not located precisely
on the Fermi surface.
Note, however, that according to Luttinger's theorem\cite{Luttinger60}
the volume enclosed by the Fermi surface is not changed
by the interaction, so that the radius of a spherically symmetric
Fermi surface is simply $k_F = \sqrt{ 2 m \mu_0}$.
On the other hand, our reference point ${\bf{k}}^{\alpha}$
is defined by
$\epsilon_{ {\bf{k}}^{\alpha} } = \mu - \delta \mu_H $, so that
to leading order
 \begin{equation}
  | {\bf{k}}^{\alpha} | - k_F   =    
 ( \mu  - \delta \mu_H - \mu_0) / {v}_F   =  \delta \mu_F  
 / {v}_F
 \; ,
 \label{eq:muren}
 \end{equation}
were we have used the fact that
to first order in the RPA screened interaction
$\mu  = \mu_0 + \delta \mu_H + \delta \mu_F$, see
Eqs.(\ref{eq:muF},\ref{eq:muH}).
It is now easy so see that for sufficiently 
regular interactions the singularity in the momentum distribution
occurs at the non-interacting $k_F$.
Substituting Eq.(\ref{eq:Gfinal}) into the definition (\ref{eq:momdis})
we obtain 
 \begin{equation}
 n ( {\bf{k}}^{\alpha} +  {\bf{q}} ) =  
 \int d {\bf{r}} e^{- i {\bf{q}} \cdot {\bf{r}} }
 \tilde{G}^{\alpha} ( {\bf{r}} ,  0 ) e^{Q_1^{\alpha} ( {\bf{r}} , 0 ) }
 \label{eq:momdis2}
 \; .
 \end{equation}
Obviously the behavior of this function for small ${\bf{q}}$  
is determined by the long-distance behavior of the
prefactor Green's function $\tilde{G}^{\alpha} ( {\bf{r}} , 0 )$
and of the Debye-Waller factor $Q_1^{\alpha} ( {\bf{r}} , 0 )$.
In a Fermi liquid $Q_1^{\alpha} ( {\bf{r}} , 0 )$ approaches
a finite constant at large distances\cite{Kopietz97,Kopietz96},
while for a Luttinger liquid the Debye-Waller factor
exhibits logarithmic singularities, which will be discussed
in Sec.3.3.
The long-distance behavior of $\tilde{G}^{\alpha} ( {\bf{r}} , 0 )$ 
is dominated by the singularities of its Fourier transform
given in Eqs.(\ref{eq:prefG}) and (\ref{eq:G1def}).
The important point is now that for finite $\delta \mu_F$
the singularity occurs at a finite value of ${\bf{q}}$.
This is of course due to the fact that we have not
measured the momenta relative to the true Fermi surface.
It is therefore convenient to
shift the momentum ${\bf{q}}$ in Eq.(\ref{eq:G1def})
by setting
 \begin{equation}
 {\bf{q}} = {\bf{q}}^{\prime} - \delta {\bf{k}}^{\alpha}
 \; ,
 \end{equation}
where $\delta {\bf{k}}^{\alpha} $ should be chosen such that it cancels
the chemical potential shift in Eq.(\ref{eq:G1def}).
For our spherical Fermi surface this
means that 
 \begin{equation}
 \delta {\bf{k}}^{\alpha}  =  {\bf{ k}}^{\alpha}  - k_F 
 \frac{ {\bf{k}}^{\alpha} }{ | {\bf{k}}^{\alpha} | } 
 = 
\frac{ \delta \mu_F }{ {v}_F}
 \frac{ {\bf{k}}^{\alpha} }{ | {\bf{k}}^{\alpha} | } 
\label{eq:pshift}
\; .
 \end{equation}
Renaming again ${\bf{q}}^{\prime} \rightarrow {\bf{q}}$,
we obtain
 \begin{equation}
 \tilde{G}^{\alpha} ( {\bf{r}} , 0 ) = e^{- i \delta {\bf{k}}^{\alpha}
 \cdot {\bf{r}}  }
 {G}^{\alpha}_2 ( {\bf{r}} , 0 ) 
 \label{eq:Gbardef}
 \; ,
 \end{equation}
with
 \begin{eqnarray}
 {G}^{\alpha}_2 ( {\bf{r}} , 0 ) 
 & =  &
 \frac{1}{\beta V}\sum_{ {\bf{q}}n }
 e^{i  {\bf{q}} \cdot {\bf{r}}  }
 \nonumber
 \\
 &  & \hspace{-10mm} \times
 \frac{
  1 + Y^{\alpha} ( {\bf{q}} 
  , i \tilde{\omega}_n ) }{
 i \tilde{\omega}_n - ( {\bf{v}}^{\alpha}  - 
 \frac{\delta {\bf{k}}^{\alpha} }{m} )
 \cdot {\bf{q}}
 - \frac{ {\bf{q}}^2}{2m} 
 - \Sigma^{\alpha}_1 ( 
 {\bf{q}}  , i \tilde{\omega}_n ) }
 \; .
 \label{eq:G1shift}
 \end{eqnarray}
Since we retain only terms to first order in $f_{\rm RPA}$,
we have ignored the shift $\delta {\bf{k}}^{\alpha}$ in the
arguments of $\Sigma_1^{\alpha}$ and $Y^{\alpha}$.
This is justified, because
according to Eqs.(\ref{eq:muF},\ref{eq:pshift}) 
the shift $\delta {\bf{k}}^{\alpha}$
is already of order $f_{\rm RPA}$.
The momentum distribution can then be written as
 \begin{equation}
 n ( {\bf{k}}^{\alpha} - \delta {\bf{k}}^{\alpha} +  {\bf{q}} ) =  
 \int d {\bf{r}} e^{- i {\bf{q}} \cdot {\bf{r}} }
 {G}^{\alpha}_2 ( {\bf{r}} ,  0 ) e^{Q_1^{\alpha} ( {\bf{r}} , 0 ) }
 \label{eq:momdis3}
 \; .
 \end{equation}
The important point is now that for a conventional
Fermi liquid 
the right-hand side of Eq.(\ref{eq:momdis3})
is non-analytic at ${\bf{q}} = 0$.
Because by construction $| {\bf{k}}^{\alpha} - \delta {\bf{k}}^{\alpha} |
= k_F = \sqrt{ 2 m \mu_0}$, the momentum
distribution is non-analytic at the non-interacting
Fermi surface, in agreement with Luttinger's theorem\cite{Luttinger60}.
In Sec.3 we shall explicitly evaluate the above
expression in one dimension, and verify that even in this case
Luttinger's theorem remains valid\cite{Blagoev97}.

The reader may find our  
discussion in this section  rather pedantic. However,
as will become evident in Sec.3,
for the calculation of the effect of the
quadratic term in the energy dispersion on the spectral
function of the TLM 
it is crucial to take the shift of  the
chemical potential due to the interaction into account.
Of course, if the location of the true Fermi surface is known
a priori
(this is the case, for example, 
in systems where Luttinger's theorem is valid and the Fermi surface
is known to be spherically symmetric),
it seems more natural to choose 
a point on the true Fermi surface as
the reference point for the linearization
of the energy dispersion.
However, for  general interactions and energy dispersions the
location of the Fermi surface is not known a priori.
In this case the general method outlined above can be used
to calculate the true Fermi surface perturbatively.
In the following section we shall 
explicitly demonstrate how this method works in one dimension.

\section{Non-linear energy dispersion in one dimension}
\label{sec:spec1d}
\noindent
In $d=1$ there are only two Fermi points, which can
be labelled by $\alpha = \pm$.
The zeroth order Green's function 
defined in Eq.(\ref{eq:G0def}) can then be written as
 \begin{equation}
 G^{\alpha}_0 ( q , i \tilde{\omega}_n )
  =  \frac{1}{
  i \tilde{\omega}_{n} -   \alpha \tilde{v}_F q 
  - \frac{ {{q}}^2}{2m} }
 \;  .
 \label{eq:energyexpansion}
 \end{equation}
The reference points $k^{\alpha}$ and the renormalized Fermi velocity
$\tilde{v}_F$ are (see Eq.(\ref{eq:kalphadef}))
 \begin{eqnarray}
 {k}^{\alpha}    & = & 
 \alpha \sqrt{ 2 m ( \mu_0 + \delta \mu_F ) } 
 \label{eq:tildekf}
 \; ,
 \\
 \tilde{v}_F & = &
 \sqrt{ 2  ( \mu_0 + \delta \mu_F ) /m } 
 \; .
 \label{eq:tildevf}
 \end{eqnarray}
It is important to emphasize that  according to the
generalized Luttinger theorem\cite{Blagoev97}
even in $d=1$
the volume of the Fermi surface  
is invariant as we switch 
on the interaction at constant density.
Hence, the Fermi wave-vector $k_F$
is still given in terms
of the chemical potential of the non-interacting system,
$k_F = \sqrt{ 2 m \mu_0}$. 
Therefore the velocity $\tilde{v}_F = | k^{\alpha} | /m$
is not identical with the 
bare Fermi velocity $v_F = k_F / m$.

In this section we shall explicitly evaluate the expressions
of Sec.2
in one dimension for a bare interaction of the form
 \begin{equation}
 f_q = f_0 \Theta ( q_c - | q | ) \; \; , \; \; q_c \ll k_F
 \; .
 \end{equation}
Note that this 
corresponds to the TLM with
forward scattering interactions $g_2 ( q ) = g_4 ( q ) = f_q$.
In one dimension the polarization function in Eq.(\ref{eq:Pol})
can for $V \rightarrow \infty$ and $\beta \rightarrow \infty$
be written as
 \begin{equation}
  \Pi_0 ( q , i \omega ) = \frac{m}{ 2 \pi q}
  \ln \left[ \frac{  ( \tilde{v}_F q + \frac{q^2}{2m} ) + \omega^2 }{
  ( \tilde{v}_F q - \frac{q^2}{2m} ) + \omega^2 } \right]
  \; .
  \label{eq:Pi1}
  \end{equation}
In the limit $1/m \rightarrow 0$ this reduces to
 \begin{equation}
  \Pi_0^{\rm lin} ( q , i \omega ) = \frac{1}{  \pi v_F}
  \frac{ ( v_F q )^2}{ (v_F q )^2 + \omega^2}
  \; \; , \; \; 1/m \rightarrow 0
  \; .
  \label{eq:PiTLM}
  \end{equation}
For finite $m$ the RPA interaction can be written as
 \begin{equation}
 f_{\rm RPA} ( q , i \omega ) = 
  \frac{ \pi \tilde{v}_F \tilde F_q}{ 1 + \tilde F_q P ( \frac{q}{2 m \tilde{v}_F} , 
 \frac{ i \omega}{ \tilde{v}_F q }) }
 \; ,
 \label{eq:frpa1d}
 \end{equation}
where we have introduced the dimensionless interaction
 \begin{equation}
 \tilde F_q = \tilde F_0 \Theta ( q_c - | q | ) \; \;  , \; \;  \tilde F_0 = \frac{f_0}{\pi \tilde{v}_F}
 \; ,
 \label{eq:Fqdef}
 \end{equation}
and the dimensionless polarization
 \begin{equation}
 P ( k , i u  )  = \frac{1}{4 k} \ln \left[ \frac{ ( 1 + {k} )^2 + u^2 }{
 ( 1 - k )^2 + u^2 } \right]
 \; .
 \label{eq:Adef}
 \end{equation}

\subsection{Chemical potential}
\noindent
Let us first consider the Fock-renormalization of the chemical potential
given in Eq.(\ref{eq:muF}), which turns out to
be finite even in $d=1$.
Introducing the dimensionless variables
$p = q / q_c$, $u = \omega / ( \tilde{v}_F q )$
and the small parameter
 \begin{equation}
 \tilde \lambda = \frac{ q_c }{ 2  m \tilde{v}_F }
 \;,\label{eq:lambda}
 \end{equation}
we obtain from Eq.(\ref{eq:muF})
 \begin{equation}
 \delta \mu_F = \langle D_0 \rangle
 = - \frac{ q_c^2}{8 \pi m } \int_{- 1}^{1}
 d p  | p | \int_{- \infty}^{\infty} d u 
 \frac{ \tilde F_0 A ( \tilde \lambda p , iu )}{ 1 + \tilde F_0 P (  \tilde \lambda p , i u ) }
 \; ,
 \label{eq:muF2}
 \end{equation}
where 
 \begin{equation}
 A ( k , i u ) = \frac{1}{ [ iu - \alpha - {k} ]
 [ iu - \alpha + k ]}
 \; .
 \label{eq:Bdef}
 \end{equation}
For small $\tilde \lambda$
we may expand the functions $P ( \tilde \lambda p , iu )$ and
$A ( \tilde \lambda p , iu )$ in powers of $\tilde \lambda$.
To leading order we obtain
 \begin{eqnarray}
 P ( \tilde \lambda p , iu ) & \approx & \frac{1}{ 1 + u^2}
 \; ,
 \label{eq:TildePi0def}
 \\
 A ( \tilde \lambda p , iu ) &  \approx &  \frac{1}{ ( i u - \alpha )^2}
 \; .
 \label{eq:A0def}
 \end{eqnarray}
Recall that $\alpha = \pm 1$ labels the two Fermi points.
The integrations in Eq.(\ref{eq:muF2})
are then easily performed, and we obtain
 \begin{equation}
  \delta \mu_F =  \frac{ q_c^2 }{4 m }  \tilde{\gamma}
  =  \frac{\tilde{v}_F q_c \tilde{\lambda}}{2}  \tilde{\gamma}
 \; ,
 \label{eq:muFres}
 \end{equation}
where
 \begin{equation}
 \tilde \gamma = \frac{ \tilde F_0^2}{2 \sqrt{1 + \tilde F_0} (
   \sqrt{1+ \tilde F_0} + 1
   )^2}
 \label{eq:anomal}
 \end{equation}
will be identified below with the
anomalous dimension of our model.
Using $\rho_0 = m \tilde{v}_F / \pi$ the Hartree 
renormalization of the chemical potential 
can be written as
 \begin{equation}
 \delta \mu_H =
  \langle i \phi_0 \rangle = 
  m \tilde{v}_F^2  \frac{\tilde F_0}{1 + \tilde F_0}
  \; ,
  \label{eq:muHres}
  \end{equation}
so that the total shift of the chemical potential is 
 \begin{equation}
 \delta \mu = \delta \mu_H + \delta \mu_F =
  m \tilde{v}_F^2  \frac{\tilde F_0}{1 + \tilde F_0}
  +  \frac{ q^2_c}{4m} \tilde \gamma
  \; .
  \label{eq:mutot1d}
  \end{equation}
Note that the Hartree renormalization is divergent in the
Tomonaga-Luttinger limit $m \rightarrow \infty$ with
$k_F / m  = {\rm const}$.
This is the reason  why it is necessary to include this
renormalization into the definition of the zeroth order
Green's function.
Note also that by construction the
prefactor Green's function (\ref{eq:G1shift})  
depends on the
bare Fermi velocity $v_F = k_F / m$.
This follows directly from
Eqs.(\ref{eq:pshift},\ref{eq:tildevf},\ref{eq:muFres}), which imply
in $d=1$
for the renormalized Fermi velocity
\begin{equation}
\tilde v_F = v_F [ 1 + \lambda^2 \tilde \gamma ] \;,
\label{eq:tildevfres}
\end{equation}
and hence for the relevant velocity that appears
in the prefactor Green's function
 \begin{equation}
 v^{\alpha}  - \frac{\delta k^{\alpha}}{m}
  =   \alpha \tilde{v}_F - \alpha \frac{ \delta \mu_F}{m \tilde{v}_F} 
  = \alpha v_F
 \; .
 \label{eq:deltavres}
 \end{equation}
Here
 \begin{equation}
 \lambda = \frac{ q_c}{2 m v_F}
\end{equation}
is defined in terms of  the bare Fermi velocity.
Obviously 
\begin{equation}
\tilde
\lambda = \lambda + O (\lambda^3)
\; , 
\end{equation}
so that to leading order in an expansion
in powers of $\lambda$ we may replace
$\tilde{\lambda} \rightarrow \lambda$.

\subsection{Prefactor Green's function}
\noindent
Next, we calculate the functions
$\Sigma_1^{\alpha} ( q , i \omega ) $ and $Y^{\alpha} ( q , i \omega )$
defined in Eqs.(\ref{eq:sigma1}) and (\ref{eq:Y})
for $V \rightarrow \infty$ and $\beta \rightarrow \infty$.
For small values of $q$ and $\omega$ and to leading
order in $f_{\rm RPA}$ it  is sufficient
to 
replace $G_1^{\alpha} \rightarrow G_0^{\alpha}$
on the right-hand sides of Eqs.(\ref{eq:sigma1}) and (\ref{eq:Y}).
This neglect of self-consistency is justified a posteriori:
because the result of the lowest order Born-approximation 
is finite and small, we presume
that the self-consistent Born-approximation is not
necessary.
Introducing the same dimensionless variables as above, we obtain
 \begin{eqnarray}
 \Sigma_1^{\alpha} ( q , i \omega )
 & = & - \tilde{v}_F q \frac{ \tilde \lambda^2}{  \pi}  
 \int_{-1}^{1} dp | p | \int_{- \infty}^{\infty} du
 \frac{ \tilde F_0 A ( \tilde \lambda p , iu )}{ 1 + \tilde F_0 P ( \tilde \lambda p , i u ) }
 \nonumber
 \\
 & \times &
 \frac{ p + \frac{q}{q_c} }{ \tilde{G}_0^{-1} + p [ i u - 
 \alpha - \tilde \lambda  ( 2 \frac{q}{q_c} + p ) ]  }
 \; ,
 \label{eq:sigma1b}
 \end{eqnarray}
 \begin{eqnarray}
 Y^{\alpha} ( q , i \omega )
 & = &  \frac{ \tilde \lambda}{  \pi} 
 \int_{-1}^{1} dp \int_{- \infty}^{\infty} du
 \frac{ \tilde F_0 A ( \tilde \lambda p , iu )}{ 1 + \tilde F_0 P ( \tilde \lambda p , i u ) }
 \nonumber
 \\
 & \times &
 \frac{ | p | + 2 {\rm sgn} (p) \frac{q}{q_c} }{ \tilde{G}_0^{-1} + p [ i u - 
 \alpha - \tilde \lambda  ( 2 \frac{q}{q_c} + {p} ) ]  }
 \; ,
 \label{eq:Yb}
 \end{eqnarray}
where 
 \begin{equation}
 \tilde{G}_0^{-1} = \frac{ i \omega - \alpha \tilde{v}_F q - 
 \frac{q^2}{2m} }{\tilde{v}_F q_c }
 \; .
 \label{eq:G0definv}
 \end{equation}
To make further progress, we expand Eqs.(\ref{eq:sigma1b}) and
(\ref{eq:Yb}) in powers of $\tilde{\lambda}$.
To leading order, we may simply set $\tilde{\lambda} = 0$ in 
the integrands and replace $\tilde v_F \to v_F$, $\tilde \lambda \to
\lambda$.
The integrations can then be performed analytically,
but the result is rather complicated and not very 
illuminating\cite{Busche99}.
For our purpose, we only need the leading terms for
small $q$ and $\omega$, which are
 \begin{equation}
 \Sigma_1^{\alpha} ( q , i \omega ) = \delta v_1^{\alpha} q
 + O (  q^2 , \omega q , \omega^2 ) + O ( \lambda^3 )
 \; \label{eq:sigma1b2},
 \end{equation}
 \begin{equation}
  \delta v_1^{\alpha}  =   - \alpha v_F \lambda^2 \gamma  
  \frac{2}{ \sqrt{1 + F_0} + 1}
  \; ,
  \label{eq:deltav1res}
  \end{equation}
where $\gamma$ is the anomalous dimension of the TLM, which is obtained
from Eq.(\ref{eq:anomal}) by replacing 
$\tilde{F}_0  \to F_0 \equiv f_0 / ( \pi v_F)$.
For the function $Y^{\alpha}_1 ( q , i \omega )$ we obtain to leading order
 \begin{equation}
 Y^{\alpha} ( q , i \omega ) = \lambda \left[ c_1  \frac{i \omega }{{v}_F q_c}
 + c_2 \frac{q}{q_c} \right] \ln \left( \frac{q_c}{ | q | } \right) +
 O ( \omega , q )
 \; ,
 \label{eq:Yres}
 \end{equation}
where $c_1$ and $c_2$ are numerical constants that depend on $F_0$.

\subsection{Debye-Waller factor}
\noindent
Finally, consider the Debye-Waller factor defined in
Eq.(\ref{eq:DebyeWaller}). 
Introducing again the dimensionless integration variables
$p = q / q_c$ and $u = \omega / ( \tilde{v}_F q )$, we obtain in
$d=1$
 \begin{eqnarray}
 Q_1^{\alpha} ( x , \tau )
 & = & \frac{1}{4 \pi} \int_{-1}^{1} \frac{dp}{| p | }
 \int_{- \infty}^{\infty} du
 \frac{ \tilde F_0 A ( \tilde \lambda p , iu )}{ 1 + \tilde F_0 P ( \tilde \lambda p , i u ) }
 \nonumber
 \\
 & \times &
 \left[ 1 - \cos ( p ( \tilde{x} - u \tilde{\tau} ) ) \right]
 \; ,
 \label{eq:Q11def}
 \end{eqnarray}
where $\tilde{x} = q_c x$ and $\tilde{\tau } = \tilde{v}_F q_c \tau $.
Since we are only interested in the leading
behavior of 
 $Q_1^{\alpha} ( x , \tau )$ for large $x$ or $\tau$, we may set
 $\tilde \lambda = 0$ in the integrand of Eq.(\ref{eq:Q11def}).
This is easily seen from the fact that 
the factor of $1/| p|$ in the integrand
is responsible for a logarithmic growth of
 $Q_1^{\alpha} ( x , \tau )$ for large $x$ or $\tau$. 
If we now expand the integrand of Eq.(\ref{eq:Q11def}) in powers
of $\tilde \lambda$, we obtain additional powers of $p$, 
so that the resulting contributions  to 
$Q_1^{\alpha} ( x , \tau )$  are bounded and hence
can be neglected as far as the
leading logarithmic  behavior is concerned.
In the limit $|\tau| \gg (\tilde
v_Fq_c)^{-1}$ and $|x| \gg q^{-1}_c$\,, the leading term is
 \begin{equation}
 Q_1^{\alpha} ( x , \tau ) \sim
 \frac{\tilde \gamma}{2}
 \ln \left[ \frac{ q_c^{-2}}{ x^2 + ( v_c \tau )^2 } \right]
 + \ln \left[ \frac{ x + i \alpha \tilde{v}_F \tau }{ 
 x + i \alpha v_c \tau } \right]
 \; ,
 \label{eq:Q1res}
 \end{equation}
where
 \begin{equation}
 v_c = \tilde{v}_F \sqrt{ 1 + \tilde F_0}
 \label{eq:vcdef}
 \end{equation}
is the velocity of collective charge oscillations and $\tilde \gamma$
is the renormalized anomalous dimension, see Eq.(\ref{eq:anomal}). 
Eq.(\ref{eq:Q1res}) 
the well-known result of the TLM, but with renormalized parameters
$\tilde \gamma, v_c$ and $\tilde v_F$. 
A more rigorous way to obtain Eq.(\ref{eq:Q1res}) 
is to first perform the $u$-integration in
Eq.(\ref{eq:Q11def}) without expanding the integrand in powers of
$\tilde \lambda$. Then, in the same limit as above, namely for large
$x$ and $\tau$, we can also
perform the $p$-integration analytically, which leads precisely to the
presented result\cite{Busche99}. Note that
Eq.(\ref{eq:Q1res}) implicitly depends on our small
parameter $\lambda$, because the renormalized 
parameters $\tilde{\gamma}$, $v_c$, and $\tilde{v}_F$ 
are $\lambda$-dependent.
Hence, Eq.(\ref{eq:Q1res}) goes beyond a simple expansion
in powers of $\lambda$.

\subsection{Spectral function}
\noindent
Combining all the results of the previous subsections,
we finally obtain for the Matsubara Green's function
for wave-vectors $q$  close to the two
Fermi points $\pm k_F = \pm \sqrt{ 2 m \mu_0}$,
 \begin{eqnarray}
 G ( \alpha k_F + q , i \omega )  & = &
 \int_{ - \infty}^{\infty} d x \int_{ - \infty }^{\infty} d \tau 
 e^{ - i ( q x - \omega \tau )}
 {G}^{\alpha}_2 ( x , \tau ) 
 \nonumber
 \\
 & \times &
 \frac{ x + i \alpha \tilde{v}_F \tau }{
 x + i \alpha v_c \tau }
 \left[ \frac{q_c^{-2}}{ x^2 + ( v_c \tau )^2 } \right]^{\tilde \gamma/2}
 \; ,
 \label{eq:Gtotres}
 \end{eqnarray}
with
 \begin{equation}
 {G}^{\alpha}_2 ( x , \tau ) 
  =  \int_{- \infty}^{\infty} \frac{dq}{2 \pi}
 \int_{- \infty}^{\infty} \frac{d\omega }{2 \pi}
 e^{  i ( q x - \omega \tau )}
 \frac{ 1 +  Y^{\alpha} ( q , i \omega )
 }{
 i \omega - \alpha {v}_1 q\, [ 1 + \lambda \frac{v_Fq}{v_1q_c}] }
 \; .
 \label{eq:prefres}
 \end{equation}
Here
 \begin{equation}
 {v}_1 = v_F 
 \left[ 1 - 
 \lambda^2 \gamma \frac{2}{ \sqrt{1 + F_0} + 1 }
 \right] + O( \lambda^3 )
 \; ,
 \label{eq:barvf}
 \end{equation}
where we have used Eqs.(\ref{eq:deltavres},\ref{eq:deltav1res}).
Because the anomalous dimension in Eq.(\ref{eq:Q1res}) 
depends on the effective dimensionless coupling
$\tilde F_0 = f_0/ ( \pi \tilde{v}_F )$, which involves the renormalized Fermi
velocity $\tilde v_F$\,,
we obtain a small correction
to the anomalous dimension $\gamma$ of the
TLM.
Keeping in mind that our result for the renormalization of the
chemical potential is based on a weak coupling expansion, 
we obtain from Eq.(\ref{eq:Q1res}) in the Tomonaga-Luttinger limit
$\lambda \rightarrow 0$ and for small $\tilde F_0$,
 \begin{equation}
 \gamma = \frac{ f_0^2 }{  8 ( \pi v_F )^2  }
 \label{gamma0}
 \; .
 \end{equation}
Hence, for finite $\lambda$ the effective 
anomalous dimension is
 \begin{equation}
 \tilde \gamma = \frac{\gamma}{ 1 + 2 \lambda^2 \gamma}
 \approx { \gamma} [ 1 
 - 2 \lambda^2 \gamma + O ( \lambda^3 ) ]
 \; .
 \label{eq:anomalres}
 \end{equation}
This is the 
leading correction to the anomalous dimension
of the TLM due to the quadratic term of the energy dispersion
at weak coupling. We expect that the non-Gaussian
corrections to the Debye-Waller factor\cite{Kopietz97}
do not change the correction term in Eq.(\ref{eq:anomalres}):
simple power counting shows that
these non-Gaussian corrections do not 
contain any infrared divergences, so that
they cannot give rise to logarithmically
growing corrections to the Debye-Waller factor, which are
necessary to renormalize the anomalous dimension.

Because our result for the vertex function $Y^{\alpha} ( q , i \omega )$
in Eq.(\ref{eq:Yres}) vanishes in the limit 
$q, \omega \to 0$, it does not modify the  qualitative behavior of the
spectral function for small $\lambda$ (Ref.\cite{Busche99}). 
To leading order, we may therefore
set $ Y^{\alpha} ( q , i \omega ) \rightarrow  0$. 
Restricting ourselves to the limit of large $x$ and $\tau$ and
neglecting the term of 
order $q^2$ in the denominator of
Eq.(\ref{eq:prefres}), we obtain 
 \begin{equation}
 {G}^{\alpha}_2 ( x , \tau )  = \frac{1}{ 2 \pi i }
 \frac{ 1}{x + i \alpha v_1 \tau }
 \; \; , \; \; |x|,|\tau| \rightarrow \infty
 \; .
 \label{eq:Gbarlin}
 \end{equation}
The imaginary-time Green's function in Eq.(\ref{eq:Gtotres}) 
becomes then 
 \begin{eqnarray}
 G ( \alpha k_F + q , i \omega )  & = &
 \frac{1}{2 \pi i}
 \int_{ - \infty}^{\infty} d x \int_{ - \infty }^{\infty} d \tau 
 e^{ - i ( q x - \omega \tau )}
 \nonumber
 \\
 &  & \hspace{-20mm} \times
 \frac{ 1 }{
 x + i \alpha v_c \tau }
 \left[ \frac{q_c^{-2}}{ x^2 + ( v_c \tau )^2 }
 \right]^{\tilde \gamma/2}\frac{x+i\alpha \tilde v_F \tau}{x+i\alpha v_1 \tau}
 \; .
 \label{eq:Gtotreslin}
 \end{eqnarray}
Note that for finite $\lambda$ and $F_0$ all three velocities
$v_1$, $\tilde{v}_F$ and $v_c$ are different.
Only for linearized energy dispersion
(i.e. for the TLM)
$\tilde{v}_F = v_1$, so that the last factor
in Eq.(\ref{eq:Gtotreslin}) is exactly unity.
For finite $\lambda$, however, the
velocity degeneracy is broken,
which has rather spectacular consequences for
the spectral function: 
there appears an additional algebraic singularity in the spectral function,
which is not present in the TLM\cite{Busche99}.
The fact that even a small breaking of the velocity degeneracy
in a special model of interacting fermions in $d=1$
can lead to new singularities in the spectral
function has recently been pointed out by
Ho and Coleman\cite{Ho99}.
Because in our model the curvature of the energy band
implies a velocity dispersion,
it is not unreasonable to expect
that the curvature
leads to new features in the spectral function, which are
not present in the TLM.
However, at this point we cannot exclude the possibility that
this new feature, whose appearance 
depends in a rather subtle way on a small renormalization
of effective velocities, is completely washed out 
by non-Gaussian corrections or by the subleading 
corrections to the Debye-Waller factor,
which have been neglected in our 
leading order calculation.
Moreover, this
singularity has a very small weight
(which can be shown\cite{Busche99}
to vanishes as $\lambda^2$ for $\lambda \rightarrow 0$), 
so that it is not of  any practical
importance. 

For these reasons we shall 
replace the last factor of Eq.(\ref{eq:Gtotreslin}) by unity
in the following analysis.
Note, however,
that the charge velocity $v_c$  
and the anomalous dimension $\tilde \gamma$ are defined in terms
of the 
renormalized velocity $\tilde{v}_F = v_F ( 1 + \lambda^2 \tilde \gamma )$,
which is slightly larger than the bare 
Fermi velocity $v_F = k_F / m$. 
To calculate the spectral function
 \begin{equation}
 A^{\alpha} ( q , \omega ) = - \frac{1}{\pi}
 {\rm Im} G  ( \alpha k_F + q , \omega + i 0^{+} )
 \label{eq:specdef}
 \end{equation}
from Eq.(\ref{eq:Gtotreslin})
with $\tilde v_F = v_1$, it is convenient to first consider the real 
time Green's function\cite{Kadanoff62} 
 \begin{equation}
 G^{\alpha}_{>} ( x , t ) 
 = i \Theta ( t ) G^{\alpha} ( x , i \tau \rightarrow - t + i 0 )
 \; .
 \end{equation}
The spectral function for $ \omega > 0$
is then  given by
 \begin{equation}
 G^{\alpha}_{>} ( q , \omega )  = 2 \pi \Theta ( \omega )
 A^{\alpha} ( q , \omega ) 
 \; ,
 \end{equation}
where
 \begin{equation}
 G^{\alpha}_{>} ( q , \omega ) 
 = i \int_{- \infty}^{\infty} dt 
 \int_{- \infty}^{\infty} dx e^{- i ( q x - \omega t )}
 G^{\alpha}_{>} ( x , t ) 
 \; .
 \end{equation}
We obtain\cite{Busche99}
 \begin{eqnarray}
 A^{\alpha} ( q , \omega )  & = &
 \left( \frac{1}{2 v_c q_c} \right)^{\tilde \gamma}
 \frac{ \tilde \gamma}{ 2 \Gamma^2 ( 1 + \frac{\tilde \gamma}{2} ) }
 \nonumber
 \\
 & \times &
 \Theta ( \omega - | v_c q | )
 ( \omega + \alpha  v_c q )^{ \frac{\tilde \gamma }{2} } 
 ( \omega - \alpha v_c q )^{ \frac{\tilde \gamma}{2} - 1}
 \; .
 \label{eq:specres}
 \end{eqnarray}
Note that $\tilde \gamma$ 
can be expanded in powers of $\lambda$, 
but a perturbative calculation of 
$A^{\alpha} ( q , \omega )$
in powers of $\lambda$ would lead to
unphysical logarithmic singularities.
To illustrate this point, consider 
the momentum distribution. From Eq.(\ref{eq:momdis2})
it is easy to show that
 \begin{equation}
 n ( \alpha k_F + q ) = \frac{1}{2}  - C \alpha 
 {\rm sgn} ( q )
  \left| \frac{q}{q_c} \right|^{\tilde \gamma} 
 \label{eq:momdis1}
 \; ,
 \end{equation}
where $C$ is a numerical constant.
Using $\tilde \gamma = \gamma - 2 \lambda^2 \gamma^2$,
an  expansion of the second term in powers of $\lambda$ 
yields
 \begin{equation}
  \left| \frac{q}{q_c} \right|^{\tilde \gamma}  =
  \left| \frac{q}{q_c} \right|^{\gamma} 
  \left[ 1 + 2 \lambda^2 \gamma \ln \left| \frac{q_c }{ q } \right| +
  O ( \lambda^4 ) \right]
  \; .
  \end{equation}
Thus, a naive expansion of $n ( \alpha k_F +  q )$ in powers
of $\lambda$ would generate logarithmic terms, which
become arbitrary large for sufficiently small $q$.
Our method effectively resums all these corrections.

\section{Conclusion} 
\noindent
In this work we have studied how
in one dimension the
quadratic term
in the expansion of the energy dispersion close to
the Fermi points modifies the spectral
function of the TLM. 
The most important effect of the non-linearity 
is a renormalization of the anomalous dimension, which
we have explicitly calculated to leading order
in the small dimensionless parameter
$\lambda = q_c / (2m v_F )$.
Note that
this implies that
correlation functions can in  general not be expanded in powers of the
$\lambda$, because
they exhibit algebraic singularities. This
leads to a non-analytic $\lambda$-dependence.
We have also pointed out that the non-linear terms in the energy dispersion
might give rise
to new features in the spectral function, which
are not present in the spectral function of the TLM.
However, we cannot exclude the possibility
that these non-universal features are 
washed out by  higher order corrections
which have been neglected in our calculation.

For a proper treatment of the quadratic term in 
the energy dispersion it is crucial to
take the  renormalization  of the chemical
potential due to the interaction into account.
We have shown how to include this effect
into the functional bosonization approach\cite{Kopietz97} 
in arbitrary dimension, emphasizing 
that the ''tadpole'' diagrams describing 
a Hartree-renormalization 
of the chemical potential have to be treated 
exactly.

\nonumsection{Acknowledgments}
\noindent
We  thank  Kurt Sch\"{o}nhammer and Rudolf Haussmann
for discussions and comments.
PK was financially supported by the
DFG via the Heisenberg program.

\nonumsection{References}
\noindent


\begin{thebibliography}{99}
\bibitem{footnote1}
We measure the frequency $\omega$ relative to  the 
chemical potential $\mu$, and 
use units such that $\hbar = 1$.
\end{thebibliography}
\end{document}